\documentclass[twocolumn,aps]{revtex4}%
\usepackage{graphicx}
\usepackage{epstopdf}

\begin{document}

\title{Pnicogen-bridged antiferromagnetic superexchange interactions in iron
pnictides}

\author{Zhong-Yi Lu$^{1}$}\email{zlu@ruc.edu.cn}
\author{Fengjie Ma$^{1,3}$}
\author{Tao Xiang$^{2,3}$}\email{txiang@aphy.iphy.ac.cn}

\affiliation{$^{1}$Department of Physics, Renmin University of
China, Beijing 100872, China}

\affiliation{$^{2}$Institute of Physics, Chinese Academy of
Sciences, Beijing 100190, China }

\affiliation{$^{3}$Institute of Theoretical Physics, Chinese Academy
of Sciences, Beijing 100190, China }

\date{June 30, 2010}

\begin{abstract}

The first-principles electronic structure calculations made
substantial contribution to study of high $T_c$ iron-pnictide
superconductors. By the calculations, LaFeAsO was first predicted to
be an antiferromagnetic semimetal, and the novel bi-collinear
antiferromagnetic order was predicted for $\alpha$-FeTe. Moreover,
based on the calculations the underlying mechanism was proposed to
be Arsenic-bridged antiferromagnetic superexchange interaction
between the next-nearest neighbor Fe moments. In this article, this
physical picture is further presented and discussed in association
with the elaborate first-principles calculations on LaFePO. The
further discussion of origin of the magnetism in iron-pnictides and
in connection with superconductivity is presented as well.

\end{abstract}


\maketitle

\section{Introduction}

Since the discovery of superconductivity in LaFeAsO by partial
substitution of O with F atoms below 26K\cite{kamihara}, intense
studies have been devoted to physical properties of iron-based
pnictides both experimentally and theoretically. Unlike study of the
cuprates, the first-principles electronic structure calculations
made substantial contribution to study of the pnictides from the
very beginning. By the calculations, we first predicted that LaFeAsO
is an antiferromagnetic semimetal \cite{ma0}, and then predicted
that the ground state of $\alpha$-FeTe is in a novel bi-collinear
antiferromagnetic order \cite{ma}. These predictions were confirmed
by the later neutron scattering experiments \cite{cruz,shi}.
Moreover, based on the calculations, we proposed \cite{ma1} the
fluctuating Fe local moments with the As-bridged antiferromagnetic
superexchange interaction to account for the magnetism and
tetragonal-orthorhombic structural distortion in LaFeAsO.

There are, so far, four types of iron-based compounds reported,
showing superconductivity after doping or under high pressures,
i.e., 1111-type $Re$FeAsO ($Re$ = rare earth) \cite{kamihara},
122-type $A$Fe$_2$As$_2$ ($A$=Ba, Sr, or Ca) \cite{rotter}, 111-type
$B$FeAs ($B$ = alkali metal) \cite{wang}, and 11-type tetragonal
$\alpha$-FeSe(Te) \cite{hsu}. Like cuprates, all these compounds
have a layered structure, namely they share the same structural
feature that there exist the robust tetrahedral layers where Fe
atoms are tetragonally coordinated by pnicogen or chalcogen atoms
and the superconduction pairing may happen. A universal finding is
that all these compounds are in a collinear antiferromagnetic order
below a tetragonal-orthorhombic structural transition temperature
\cite{cruz,dong} except for $\alpha$-FeTe that is in a bi-collinear
antiferromagnetic order below a tetragonal-triclinic structural
transition temperature \cite{ma,bao,shi}.

The above finding  raises the question on the microscopic mechanisms
underlying the structural transition and antiferromagnetic
transition respectively and the relationship between these two
transitions. There are basically two contradictive views upon this
question. The first one \cite{mazin} is based on itinerant electron
picture, which thinks that there are no local moments and the
collinear antiferromagnetic order is induced by the Fermi surface
nesting which is also responsible for the structural transition due
to breaking the four-fold rotational symmetry. On the contrary, the
second one is based on local moment picture. The $J_1$-$J_2$
Heisenberg model was, phenomenologically \cite{yildirim} and from
strong electron correlation limit \cite{si} respectively, proposed
to account for the issue. Meanwhile, we proposed \cite{ma1} the
fluctuating Fe local moments with the As-bridged antiferromagnetic
superexchange interactions as the driving force upon the two
transitions, effectively described by the $J_1$-$J_2$ Heisenberg
model as well.

Our proposal embodies the twofold meanings shown by the calculations
\cite{ma1}: (1) there are localized magnetic moments around Fe ions
and embedded in itinerant electrons in real space; (2) it is those
bands far from the Fermi energy rather than the bands nearby the
Fermi energy that determine the magnetic behavior of pnictides,
namely the hybridization of Fe with the neighbor As atoms plays a
substantial role. Here the formation of local moment on Fe ion is
mainly due to the strong Hund's rule coupling on Fe $3d$-orbitals
\cite{ma1}. In this sense, our proposal can be considered as the
Hund's rule correlation picture. We emphasize again that the Arsenic
atoms play a substantial role in our physical picture. Subsequently
we successfully predicted from the calculations \cite{ma1}, based on
this Hund's rule correlation picture, that the ground state of
$\alpha$-FeTe is in a bi-collinear antiferromagnetic order, which
was later confirmed by the neutron scattering experiments
\cite{shi,bao}.

Here we would like to address that our physical picture works well
on all iron-based pnictide superconductors, including LaFePO. In
this article, we will show how our physical picture works on LaFePO
through the elaborate first-principles electronic structure
calculations. We find that there is P-bridged next-nearest neighbor
Fe-Fe superexchange antiferromagnetic interaction while the nearest
neighbor Fe-Fe exchange interaction is very small in LaFePO. This
results in the collinear antiferromagnetic order of Fe moments in
the ground state, similar to the scenario in LaFeAsO. Moreover, we
demonstrate that there also exits a small monoclinic lattice
distortion resulting from this superexchange magnetic interaction,
again similar to the one in LaFeAsO \cite{ma1}.

\section{Computational Approach}

In the calculations the plane wave basis method was used
\cite{pwscf}. We used the generalized gradient approximation (GGA)
of Perdew-Burke-Ernzerhof \cite{pbe} for the exchange-correlation
potentials. The ultrasoft pseudopotentials \cite{vanderbilt} were
used to model the electron-ion interactions. After the full
convergence test, the kinetic energy cut-off and the charge density
cut-off of the plane wave basis were chosen to be 600eV and 4800eV,
respectively. The Gaussian broadening technique was used and a mesh
of $16\times 16\times 8$ k-points were sampled for the
Brillouin-zone integration. In the calculation, the internal atomic
coordinates within the cell were determined by the energy
minimization.

The crystal structure of LaFePO belongs to the tetragonal structure
and the space group is of P4/nmm. In the calculations, we adopted
two unit cells, namely $a\times a\times c$ crystal unit cell and
$\sqrt{2}a\times\sqrt{2}a\times c$ unit cell respectively, as shown
in Fig. \ref{fig:1}. Here we used the experimental lattice
parameters with a=3.964 \AA~ and c=8.512 \AA. We also checked the
results with the optimized lattice parameters (a=b=3.932 \AA,
c=8.415 \AA) and there is no significant change found.

\section{Results and analysis}

\begin{figure}
\includegraphics[width=5cm]{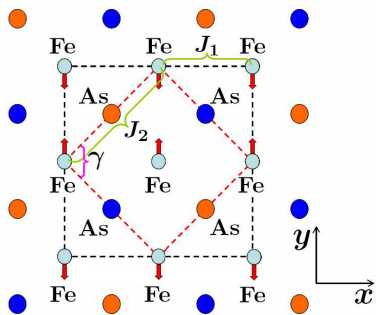}
\caption{(Color online) Schematic top view of the FeAs or FeP layer
in LaFeAsO or LaFePO. The small dashed square is an $a\times a$ unit
cell while the large dashed square is a $\sqrt{2}a\times \sqrt{2} a$
unit cell. The Fe-spins are aligned in a collinear antiferromagnetic
order.}\label{fig:1}
\end{figure}

In order to derive the exchange interactions between the Fe-Fe
moments, we designed three different magnetic structures, including
the nonmagnetic, the ferromagnetic, and the checkerboard (Neel)
antiferromagnetic structures calculated in $a\times a\times c$
crystal unit cell, and the collinear antiferromagnetic structure
calculated in $\sqrt{2}a\times \sqrt{2} a\times c$ unit cell. The
calculations show that the energy differences among the nonmagnetic,
the ferromagnetic, and the checkerboard antiferromagnetic states are
so small that these three states can be considered as degenerate,
similar to the ones in Ref. \cite{che}. The calculated electronic
band structure and the Fermi surface are also the same as the ones
in the previous calculations \cite{lebegue,che}. From the volumes
enclosed by the Fermi surface sheets, we derive that both hole and
electron carrier density is about $2.75\times 10^{21}/cm^3$. Thus
LaFePO in the nonmagnetic phase behaves as a semimetal. However, the
calculation upon the collinear antiferromagnetic structure in LaFePO
shows that the ground state of LaFePO is of the collinear
antiferromagnetic ordering on the Fe moments. The energy of the
collinear antiferromagnetic ordered state is about 33 meV per
formula unit cell lower than the nonmagnetic one. The magnetic
moment around each Fe atom is found to be nearly $2.0 \mu_{B}$,
smaller than the one in the calculation on LaFeAsO \cite{ma1}.

These magnetic states can be modeled by the following frustrated
Heisenberg model with the nearest and next-nearest neighbor
couplings $J_1$ and $J_2$
\begin{equation}\label{eq:Heisenberg}
H=J_1\sum_{\langle ij \rangle}\vec{S}_i\cdot\vec{S}_j +J_2\sum_{ \ll
ij \gg}\vec{S}_i\cdot\vec{S}_j,
\end{equation}
whereas $\langle ij \rangle$ and $\ll ij \gg$ denote the summation
over the nearest and next-nearest neighbors, respectively. From the
calculated energy data, we find that $J_1 \sim 0.5 meV/S^2$ and $J_2
\sim 7.5 meV/S^2.$ If the spin of each Fe ion $S= 1$, then $J_1 \sim
0.5 meV$ and $J_2 \sim 7.5 meV$ (The detailed calculation is
referred to Appendix of Ref. \onlinecite{ma1}). In comparison with
LaFeAsO \cite{ma1}, the $J_1$ and $J_2$ are smaller by 7 times for
LaFePO.

\begin{figure}
\includegraphics[width=7.5cm]{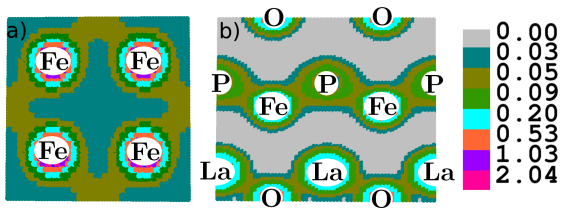}
\caption{(Color online) Calculated charge density distribution of
LaFePO in the (001) plane crossing Fe-Fe atoms (a) and in the (110)
plane crossing Fe-P-Fe atoms (b). } \label{fig:2}
\end{figure}

Fig. \ref{fig:2} (a) and (b) show the calculated charge
distributions on the (001) plane crossing four Fe atoms and the
plane (110) crossing Fe-P-Fe atoms, respectively. We see that there
is a strong bonding between Fe and P ions. This indicates that the
next-nearest neighbor coupling $J_2$ is induced by the superexchange
bridged by P $3p$-orbitals. This superexchange is antiferromagnetic
because the intermediated state associated with the virtual hopping
bridged by P ions is a spin singlet. The charge distribution between
the two nearest Fe ions is finite but almost none between the
next-nearest Fe ions. Thus there is a direct exchange interaction
between the two nearest neighboring Fe spins. Because of the exists
of the strong Hund's coupling induced direct ferromagnetic exchange
interaction and the indirect antiferromagnetic superexchange
interaction bridged by P $3p$-orbitals, combined with the
degeneration of different magnetic states in $a\times a\times c$,
the nearest neighbor coupling $J_1$ is nearly zero.

For the collinear antiferromagnetic structure, our calculations show
that there is further small structural distortion that shrinks along
the spin-parallel alignment (x-axis in Fig. 1) and expands along the
spin-antiparallel alignment (y-axis in Fig. 1), causing structural
transition from tetragonal structure to monoclinic structure. This
is easily understandable that the direct exchange favors a shorter
Fe-Fe separation while the superexchange favors a larger Fe-P-Fe
angle. It turns out that the angle $\gamma$ in ab-plane (see Fig.
\ref{fig:1}) is no longer rectangular and optimized to be
$90.57^{\circ}$ with a small energy gain of about 3meV, similar to
the finding in LaFeAsO \cite{ma1}. Such a small distortion is found
to have a small effect upon the electronic structure and the Fe
moments. Here we emphasize that our calculations show that the
driving force upon this structural distortion is nothing but the
magnetic interaction. More specifically it is the superexchange
interaction $J_2$ that breaks the rotational symmetry to induce both
the structural distortion and the Fe-spin collinear
antiferromagnetic ordering.

\begin{figure}
\includegraphics[width=7.5cm]{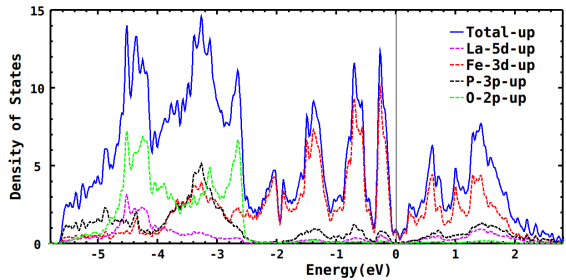}
\caption{(Color online) Calculated total and orbital-resolved
partial density of states (spin-up part) of LaFePO in the collinear
antiferromagnetic state. The Fermi energy is set to zero. }
\label{fig:3}
\end{figure}

Fig. \ref{fig:3} shows the total and projected density of states of
LaFePO in the collinear antiferromagnetic state. The most of the
states around the Fermi level are of Fe 3d characters. The
corresponding electronic density of states is 1.45 states per eV per
formula unit cell, which is less than the one of nomagnetic state
(=5.78 states per eV per formula unit cell). The specific heats are
evaluated as 0.84 mJ/(K$^2\ast$mol) and 6.86 mJ/(K$^2\ast$mol) for
the collinear antiferromagnetic state and nonmagnetic state
respectively, consistent with the specific heat measurement
\cite{mcqueen}. The paramagnetic susceptibility in the nonmagnetic
state is about $1.18\times 10^{-9} m^{3}/mol$.

\begin{figure}
\includegraphics[width=8.5cm]{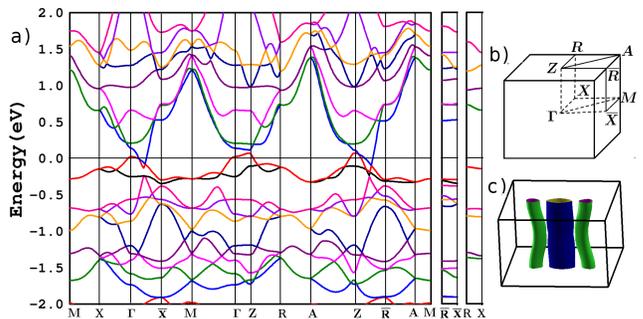}
\caption{(Color online) (a) The electronic band structure of LaFePO
in the collinear antiferromagnetic state. The Fermi energy is set to
zero. (b) The correspondent labels in Brillouin zone. (c) The Fermi
surface sheets: a hole-type cylinder along $\Gamma$Z and two
electron-type pockets between $\Gamma$ and $\bar{X}$. $\Gamma$X
corresponds to the parallel-aligned moment line and
$\Gamma$$\bar{X}$ corresponds to the antiparallel-aligned moment
line.} \label{fig:5}
\end{figure}

Fig. \ref{fig:5} shows the collinear antiferromagnetic electronic
band structure and the Fermi surface. Unlike the ones in the
nonmagnetic state \cite{lebegue}, there are now only two bands
crossing the Fermi level, one small hole cylinder along $\Gamma$Z
and two small electron cylinders formed between $\Gamma$ and
$\bar{X}$. From the volumes enclosed by these Fermi surface sheets,
we find that the hole carrier density is about
$1.34\times10^{20}/cm^3$ and the electron carrier density is about
$1.21\times10^{20}/cm^3$.

\section{Origin of the magnetism}

\begin{figure}
\includegraphics[width=7.5cm]{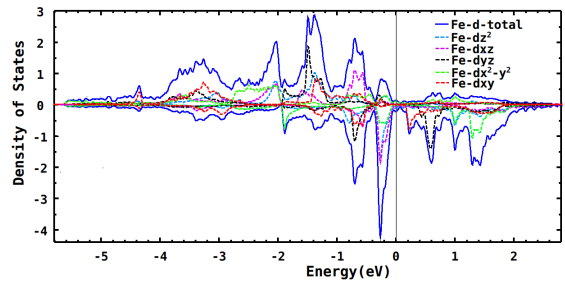}
\caption{(Color online) Calculated total and projected density of
states at the five Fe-$3d$ orbitals around one of the four Fe atoms
in the collinear antiferromagnetic state unit cell. The Fermi energy
is set to zero.}\label{fig:4}
\end{figure}

By projecting the density of states onto the five $3d$ orbitals of
Fe in the collinear antiferromagnetic state of LaFePO (Fig.
\ref{fig:4}), we find that the five up-spin orbitals are almost
filled and the five down-spin orbitals are only partially filled.
Moreover, the down-spin electrons are nearly uniformly distributed
in these five 3d orbitals. This indicates that the crystal field
splitting imposed by P atoms is relatively small and the Fe
3d-orbitals hybridize strongly with each other. As the Hund rule
coupling is strong, this would lead to a magnetic moment formed
around each Fe atom, as found in our calculation. Actually, this is
a universal feature for all iron pnictides, as we first found in
LaFeAsO \cite{ma1}. The polarization of Fe magnetic moments is thus
due to the Hund's rule coupling.

In low temperatures, the Fe moments will interact with each other to
form an antiferromagnetic ordered state. These ordered magnetic
moments have been observed by elastic neutron scattering and other
experiments in LaFeAsO and other pnictides\cite{cruz,bao}. However,
they are not exactly the moments obtained by the density functional
theory (DFT) calculations, as we indicated first for LaFeAsO
\cite{ma1}. This is because the DFT calculation is done based on a
small magnetic unit cell and the low-energy quantum spin
fluctuations as well as their interactions with itinerant electrons
are frozen by the finite excitation gap due to the finite-size
effect. Thus the moment obtained by the DFT is the bare moment of
each Fe ion. It should be larger than the ordered moment measured by
the neutron scattering and other experiments. Our calculations show
that the bare magnetic moment around each Fe atom is about $2\mu_B$.
Moreover, the frustration between the $J_1$ and $J_2$ terms will
further suppress the antiferromagnetic ordering at the two
Fe-sublattices, each connected only by the $J_2$ terms. All of these
together will reduce strongly the average magnetic ordered moment
around each Fe measured by experiments, like the neutron scattering.

In high temperatures, there is no net static moment in the
paramagnetic phase due to the thermal fluctuation, but the bare
moment of each Fe ion can still be measured by a fast local probe
like ESR (electron spin resonance). Recently the bare moment of Fe
has been observed in the paramagnetic phase of BaFe$_2$As$_2$ by the
ESR measurement \cite{chen}. The value of the moment detected by ESR
is about $2.2\sim 2.8~\mu_B$ in good agreement with our DFT result
\cite{ma1}, which is but significantly larger than the ordered
moment in the antiferromagnetic phase.

\section{Discussion in connection with superconductivity}

\begin{figure}
\includegraphics[width=8cm]{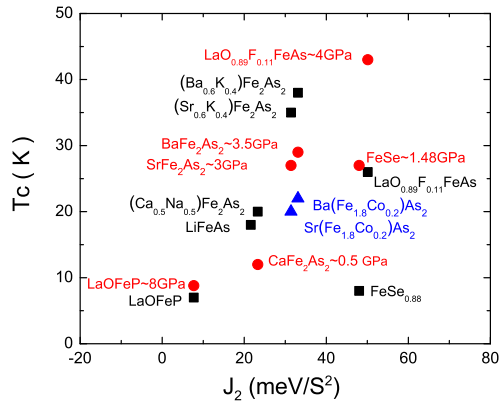}
\caption{Various critical superconductivity temperature $T_c$ versus
antiferromagnetic superexchange interaction $J_2$ for iron-based
pnictides or chalcogenides with doping or under high pressures. Tc
are taken from Refs.
\onlinecite{kamihara,rotter,wang,hsu,122P,LOFAP,BFACo,
SFACo,FeSeP,SFAK,CFANa,LOFP,LOFPP}.}\label{Tc-J2}
\end{figure}

As we see above, the next nearest neighbor antiferromagnetic
superexchange interaction affects strongly the magnetic structure of
the ground state. The magnetic fluctuation induced by the
antiferromagnetic superexchange interactions should be responsible
for the superconducting pairing. Upon doping, the long range
ordering will be suppressed. However, we believe that the remanent
antiferromagnetic fluctuation will survive, similar as in cuprate
superconductors. We plot the superconducting critical temperatures
$T_c$ versus $J_2$ for the iron pnictide compounds with doping
carriers or by applying high pressures in Fig. \ref{Tc-J2}. As we
see, the maximum critical temperatures $T_c$ are in proportion to
the next-nearest neighbor antiferromagnetic superexchange
interaction $J_2$. This suggests that there would exist an intrinsic
relationship between the superconductivity and the Pnicogen-bridged
antiferromagnetic superexchange interactions.

The above analysis suggests that there are some similarities between
pnictide and cuprate superconductors. Here we would also like to
further address the difference between them is that the undoped
cuprate is an antiferromagnetic Mott insulator while the undoped
pnictide is an antiferromagnetic semimetal. In particular, in
undoped pnictide there already coexist both local magnetic moments
and itinerant electrons. This is similar as in the doped cuprate
superconductors. In cuprate superconductors, the low-energy physics
can be effectively described by the $t-J$ model, in which the
coupling between local spins is the antiferromagnetic interaction
and the doped holes or electrons can hop on the lattice.

For iron-based pnictides, the low energy spin dynamics can be
approximately described by an antiferromagnetic Heisenberg model
(Eq. (\ref{eq:Heisenberg}) with the nearest and the next-nearest
neighbor exchange interactions. However, the Fe spin (or magnetic
moment) is not quantized since the electrons constituting the moment
can propagate on the lattice, like in hole or electron doped high
$T_c$ superconductivity cuprates. Besides these superexchange
interactions, the on-site Hund's rule coupling among different Fe
$3d$ orbitals is important. This is because the crystal splitting of
Fe $3d$ levels is very small and the spins of Fe $3d$ electrons are
polarized mainly by this interaction. Thus we believe that the
low-energy physical properties of these iron-based pnictides can be
approximately described by the following effective Hamiltonian
\begin{eqnarray}\label{eq:1}
H & = &\sum_{ \langle ij\rangle,\alpha\beta}t_{ij}^{\alpha\beta}
c_{i\alpha}^{\dagger}c_{j\beta} - J_H \sum_{i,\alpha\not=
\beta}\vec{S}_{i\alpha}\cdot\vec{S}_{i\beta}
\nonumber \\
&& + J_1\sum_{\langle ij \rangle,
\alpha\beta}\vec{S}_{i\alpha}\cdot\vec{S}_{j\beta}
+J_2\sum_{\langle\langle ij\rangle\rangle,
\alpha\beta}\vec{S}_{i\alpha} \cdot\vec{S}_{j\beta} ,
\end{eqnarray}
where $\langle ij \rangle$ and $\langle\langle ij \rangle\rangle$
represent the summation over the nearest and the next-nearest
neighbors, respectively. $\alpha$ and $\beta$ are the indices of Fe
$3d$ orbitals. $c^{\dagger}_{i\alpha}~(c_{i\alpha})$ is the electron
creation (annihilation) operator.
\begin{equation}
\vec{S}_{i\alpha} = c_{i\alpha}^\dagger \frac{\vec{\sigma}}{2}
c_{i\alpha}
\end{equation}
is the spin operator of the $\alpha$ orbital at site $i$. The total
spin operator at site $i$ is defined by
$\vec{S}_i=\sum_{\alpha}\vec{S}_{i\alpha}$. In Eq.~(\ref{eq:1}),
$J_H$ is the on-site Hund's coupling among the five Fe $3d$
orbitals. The value of $J_H$ is generally believed to be about 1 eV.
$t_{ij}^{\alpha\beta}$ are the effective hopping integrals that can
be determined from the electronic band structure in the nonmagnetic
state\cite{ccao}. It has been shown that the nearest and the
next-nearest neighbor exchange coupling constants $J_1$ and $J_2$ in
Eq.~(\ref{eq:1}) can be calculated from the relative energies of the
ferromagnetic, square antiferromagnetic, and collinear
antiferromagnetic states with respect to the non-magnetic state. For
the corresponding detailed calculations, please refer to the
appendix in our paper in Ref. \onlinecite{ma1}. Here we have assumed
that the contribution of itinerant electrons to the energy is almost
unchanged in different magnetically ordered states. Since the bare
$t_{ij}^{\alpha\beta}$ and $J_H$ can be considering independent of
magnetic structures, the relative energies between different
magnetic states are not affected by itinerant electrons.

\section{Conclusion}

In conclusion, we have presented the first-principles calculation of
the electronic structure of LaFePO. We find that like the situation
in LaFeAsO, there are the next-nearest neighbor antiferromagnetic
superexchange interactions bridged by P $3p$-orbitals, which are
much larger than the nearest neighbor ones. This gives rise to the
collinear antiferromagnetic ordering of Fe spins in the ground state
as observed in LaFeAsO by the neutron scattering. However, because
of much smaller exchange interactions $J_2$ and $J_1$ in comparison
with the ones in LaFeAsO, the corresponding transition temperatures
should be very low in LaFePO, very likely less than 10K scaled
according to LaFeAsO. This needs further experiment to test and
verify. Based on the analysis of electronic and magnetic structures,
we proposed that the low-energy physics of iron pnictides can be
effectively described by the $t-J_H-J_1-J_2$ model, defined by
Eq.~(\ref{eq:1}).

\acknowledgements

This work is supported by National Natural Science Foundation of
China and by National Program for Basic Research of MOST, China.


\begin{references}

\bibitem{kamihara}Y. Kamihara, T. Watanabe, M. Hirano, and H. Hosono,
J. Am. Chem. Soc. {\bf 130}, 3296 (2008).
\bibitem{ma0}F. Ma and Z.Y. Lu, Phys. Rev. B {\bf 78}, 033111
(2008).
\bibitem{ma} F. Ma {\it et al.}, Phys. Rev. Lett. {\bf 102},
177003 (2009).
\bibitem{cruz} C. de la Cruz {\it et al.}, Nature (London)
{\bf 453}, 899 (2008).
\bibitem{shi} S.L. Li {\it et al.}, Phys. Rev.
B {\bf 79}, 054503 (2009).
\bibitem{ma1}F. Ma, Z.Y. Lu, and T. Xiang, Phys. Rev. B {\bf 78}, 224517 (2008).
\bibitem{rotter}M. Rotter {\it et al.}, Phys. Rev. Lett. {\bf 101}, 107006 (2008).
\bibitem{wang}X. C. Wang {\it et al.}, Solid State Commun. {\bf 148}, 538 (2008).
\bibitem{hsu}F.-C. Hsu {\it et al.}, Proc. Natl. Acad. Sci. U.S.A. {\bf 105},
14262 (2008).
\bibitem{dong} J. Dong {\it et al.}, Europhys. Lett. {\bf 83}, 27006 (2008).
\bibitem{bao} W. Bao {\it et al.}, Phys. Rev. Lett. {\bf 102}, 247001 (2009).
\bibitem{mazin}I.I. Mazin, D.J. Singh, M.D. Johannes, and M.H. Du,
Phys. Rev. Lett. {\bf 101}, 057003 (2008).
\bibitem{yildirim} T. Yildirim, Phys. Rev. Lett. {\bf 101}, 057010
(2008).
\bibitem{si} Q. Si and E. Abrahams, Phys. Rev. Lett. {\bf 101}, 076401
(2008).
\bibitem{pwscf}P. Giannozzi et al., http://www.quantum-espresso.org.
\bibitem{pbe}J. P. Perdew, K. Burke, and M. Erznerhof,
Phys. Rev. Lett. {\bf 77}, 3865 (1996).
\bibitem{vanderbilt}D. Vanderbilt, Phys. Rev. B {\bf 41},
7892 (1990).
\bibitem{che}Renchao Che, Ruijuan Xiao, Chongyun Liang,
Huaixin Yang, Chao Ma, Honglong Shi, and Jianqi Li, Phys. Rev. B
{\bf 77}, 184518 (2008).
\bibitem{lebegue}S. Leb\`{e}gue, Phys. Rev. B {\bf 75}, 035110 (2007).
\bibitem{mcqueen}T. M. McQueen, M. Regulacio, A.J.
Williams, Q. Huang, J.W. Lynn, Y.S. Hor, D.V. West, M.A. Green, and
R.J. Cava, Phys. Rev. B {\bf 78}, 024521 (2008).
\bibitem{chen} T. Wu, J.J. Ying, G. Wu, R.H. Liu, Y. He, H. Chen,
X.F. Wang, Y.L. Xie, Y.J. Yan, and X.H. Chen, Phys. Rev. B {\bf 79},
115121 (2009).

\bibitem{122P}P. L. Alireza, J. Gillett,
Y. T. C. Ko, S. E. Sebastian, and G. G. Lonzarich, J. Phys.:
Condens. Matter {\bf 21}, 012208 (2009).
\bibitem{LOFAP}H. Takahashi,
K. Igawa, K. Arii, Y. Kamihara, M. Hirano, and H. Hosono, Nature
{\bf 453}, 376 (2008).
\bibitem{BFACo}A. S. Sefat, R. Jin,M. A.
McGuire, B. C. Sales, D. J. Singh, and D. Mandrus, Phys. Rev. Lett.
{\bf 101}, 117004 (2008).
\bibitem{SFACo}A. Leithe-Jasper, W. Schnelle, C. Geibel, and H. Rosner,
Phys. Rev. Lett. {\bf 101}, 207004 (2008).
\bibitem{FeSeP}Y. Mizuguchi, F. Tomioka, S. Tsuda, T. Yamaguchi, and Y.
Takano, Appl. Phys. Lett. {\bf 93}, 152505 (2008).
\bibitem{SFAK}G. F. Chen, Z. Li, J. Dong, G. Li, W. Z. Hu, X. D. Zhang, X.
H. Song, P. Zheng, N. L. Wang, and J. L. Luo, Phys. Rev. B {\bf 78},
224512 (2008).
\bibitem{CFANa} G.Wu, H. Chen, T.Wu, Y. L. Xie, Y. J. Yan, R. H. Liu, X. F.
Wang, J. J. Ying, and X. H. Chen, J. Phys.: Condens. Matter, {\bf
20}, 422201 (2008).
\bibitem{LOFP}Y. Kamihara, H. Hiramatsu, M. Hirano, R. Kawamura, H.
Yanagi, T. Kamiya, and H. Hosono, J. Am. Chem. Soc. {\bf 128}, 10012
(2006).
\bibitem{LOFPP}K. Igawa, H. Okada, K. Arii, H. Takahashi, Y. Kamihara, M.
Hirano, H. Hosono, S. Nakano, and T. Kikegawa, J. Phys. Soc. Jpn.
{\bf 78}, 023701 (2009).
\bibitem{ccao}C. Cao, P.J. Hirschfeld, and H.P. Cheng,
Phys. Rev. B {\bf 77}, 220506(R) (2008).

\end{references}
\end{document}